\pdfoutput=1
\documentclass[12pt,letterpaper,floatfix]{article}
\usepackage{amssymb}
\usepackage{hyperref}
\usepackage{cite}
\usepackage{amsmath,amssymb,amsthm,bbm,color}
\usepackage{graphicx}
\usepackage{epsfig}
\usepackage{euscript}
\usepackage{pslatex}
\usepackage{fancybox}
\usepackage{enumerate}
\usepackage{widetext}
\usepackage{subfigure}
\usepackage{wasysym}
\usepackage[compat=1.0.0]{tikz-feynman}

\newcommand{\Meu}{\EuScript{M}}

\newcommand{\KK}{${\cal KK}$}

\newcommand{\sfac}{\mathfrak{s}}

\def\st{\hbox{}} 
\usepackage[figuresright]{rotating}

\begin{document}
\begin{titlepage}

\begin{center}
\begin{flushleft}
{\small \bf BU-HEPP-23-01, Jan. 2023}
\end{flushleft}
\vspace{18mm}

{\bf Collinearly Enhanced Realizations of the Yennie-Frautschi-Suura (YFS) MC Approach to Precision Resummation Theory}\\
\vspace{2mm}

{S. Jadach\footnote{Deceased.}$^{,\;a}$,~B.F.L. Ward$^b$,~ Z. Was$^a$}\\
{$^a$Institute of Nuclear Physics, Polish Academy of Sciences, Krakow, PL}\\
{$^b$Baylor University, Waco, TX, USA}\\
\end{center}
\centerline{\bf Abstract}
We extend the Yennie-Frautschi-Suura (YFS) IR resummation theory to include all of the attendant collinear contributions which exponentiate. This improves the original YFS formulation in which only a part of these contributions was exponentiated. 
We show that the new resummed contributions agree with known results from the collinear factorization approach and we argue that they improve the attendant precision tag for a given level of exactness in the respective YFS hard radiation residuals. 
\end{titlepage}

\section{Introduction}
\label{intro}
The exact amplitude-based  CEEX/EEX YFS MC approach to EW higher order corrections is given in 
Refs.~\cite{Jadach:1993yv,Jadach:1999vf,Jadach:2000ir,Jadach:2013aha}. Here, CEEX denotes the coherent exclusive exponentiation developed in Refs.~\cite{Jadach:1999vf,Jadach:2000ir,Jadach:2013aha} in which IR singularities are resummed at the level of the amplitude. EEX denotes exclusive exponentiation as originally formulated by Yennie, Frautschi and Suura (YFS) in Ref.~\cite{yfs:1961} and it is effected at the squared amplitude level. In the context of precision physics for $e^+e^-$ colliding beam devices, we and our collaborators have developed and implemented several MC event generators which realize the YFS MC approach for EW higher order corrections -- see Ref.~\cite{frix-lnen:2022} for a concise catalog of these programs. All of MC's in that catalog, except perhaps for KORALZ~\cite{Jadach:1993yv}, which has been superseded by {\KK}MC-ee~\cite{Jadach:1999vf,dizet642,Jadach:2022vf}, are under consideration for the appropriate upgrades to meet the new precision requirements for the future $e^+e^-$ colliders: FCC-ee~\cite{Blondel:2019vdq}, CLIC~\cite{CLIC-burrows}, ILC~\cite{ILC-behnke}, and CEPC~\cite{CEPC-gao}. 

While the physics expectations for the future colliding beam devices are well-defined and formulated~\cite{Jadach:2019bye,frix-lnen:2022}, in what follows we present a collinear improvement of the original YFS algebra with an eye toward enhancing these physics expectations\footnote{The enhanced precision which results from collinear improvement of YFS theory also obtains at current $e^+e^-$ colliders since it will be seen that the size of the respective effects scales as $\frac{\alpha}{\pi}L\cong 0.0352,\; 0.0416,\; \text{and}\; 0.0463$ respectively at DAFNE, BESIII, and BELLEII vs $0.0558$ at FCC-ee$|_{M_Z}$. The latter case has the more strict precision requirements so that the enhancement is more relevant. Here, $L$ is the respective big log defined in the discussion below.}. We argue that this improvement yields even more precision for a given level of exactness in the respective hard photon residuals $\bar{\beta}_n \; (\hat{\beta}_n)$ in the EEX (CEEX) formulation respectively. This work builds on ideas presented by one of us (SJ) already in Ref.~\cite{Jadach-unpublhd-2002}. Here, we note that there are two separate but related problems to be addressed: one may start with the YFS algebra and its exactness in the infrared limit to all orders in $\alpha$ and improve its resummation of non-soft collinear big logs that are incompletely resummed but are treated to any given level of desired exactness using the hard photon residuals $\bar\beta_n(\hat\beta_n)$ as defined in Refs.~\cite{yfs:1961,Jadach:1999vf}, respectively; or, one may start with the collinear factorization approach based on DGLAP theory~\cite{dglap1,dglap2,dglap3,gribv-lptv:1972,dglap5} which treats collinear big logs to all orders but is not exact in the infrared limit and improve its resummation of the soft non-collinear infrared limit. In Ref.~\cite{Jadach-unpublhd-2002}, the latter problem was addressed. Here, the former problem is addressed.\par

We note that collinear factorization approach has recently been improved to the next-leading log level in Refs.~\cite{frixione-2019,bertone-2019}. In addition to the differences in the YFS and collinear factorization approaches just noted,  we point out another fundamental difference between the two. The exact phase space for the multiple photon radiation in $e^+e^-\rightarrow \bar{f}f+n\gamma, n>0,$ is realized on an event by event basis to all orders in 
$\alpha$ whereas in the collinear factorization approach, realizaed via structure functions, the radiation transverse degrees of freedom that have been integrated to reach the 1-dimensional structure function distribution have to be restored and this restoration is inherently approximate, as it was illustrated in Ref.~\cite{bhabhacern-lep2:1996},for example\footnote{In other words, the distributions which the structure function (collinear factorization) approach produces are not exact for the transverse degrees of freedom which were integrated out to arrive at the collinear limit represented by the structure functions while our distributions are exact in these degrees of freedom. We have seen in the LEP studies~\cite{bhabhacern-lep2:1996} that the detailed measurements of the exclusive photon distributions show this deviation from exactness.}. 
\par

Application of the structure function realized collinear factorization approach is limited to ``academic observables'' with a cut-off on the total photon energy $E_{\max} = (sx_1x_2)^{1/2}$ where the $x_i$ are the respective parton momentum fractions\footnote{In any real observable there is always multiple photon radiation to all orders in $\alpha$. Any fixed-order calculation thus is necessarily academic, and its usefulness has to determined on a case-by-case basis. In many cases, the effects of the multi-photons missing from the fixed-order result are small enough that the fixed-order result can be used to assess the data. Even in the latter cases, the lack of exactness of the treatment of the transverse degrees of freedom in the collinear factorization approach limits its applicability.}.
All realistic experimental observables select events using acollinearity
and other similar cuts depending on photon momenta in a complicated way.
Only a Monte Carlo with full multiple photon phase space can 
provide predictions for the real experiments.
On the other hand, variants of the structure function realized collinear factorization approach with added sub-leading corrections in the structure functions and the respective hard sub-process parton-parton cross sections are quite useful in testing/calibrating
Monte Carlo programs.
For instance the BHLUMI program~\cite{bhlumi2:1992,bhlumi4:1996} includes the structure function
based program LUMLOG, while \KK{MC}~\cite{Jadach:1999vf} provides the KKsem and KKfoam auxiliary programs,
which serve for testing/calibrating the main multiphoton generator,
albeit for academic observables.\par

The discussion proceeds as follows. Since it is still not generally used, we first review in the next Section  CEEX/EEX realization of the higher order EW corrections to the Standard Theory (ST)\footnote{We follow Prof. D.J. Gross~\cite{djg-smat50} and refer to the Standard Model as the Standard Theory of elementary particles.}~\cite{SM1,SM2,SM3,SM4} of elementary particles. Then we show in Section 3 how to extend the attendant YFS IR algebra to include the respective complete collinear contribution. We compare with known collinear leading log results. We close with some summary remarks in Section 4. 
\section{Brief Review of CEEX/EEX Realization of Higher Order EW Corrections}
Specifically, we recall the master formula for the CEEX/EEX realization of the YFS resummation of the EW Standard theory.
For the prototypical process 
$e^+e^-\rightarrow f\bar{f}+n\gamma, \; f = \ell, q, \; \ell=e,\mu,\tau,\nu_e,\nu_\mu,\nu_\tau, \; q = u,d,s,c,b,t$, we have the cross section formula
\begin{equation}
\sigma =\frac{1}{\text{flux}}\sum_{n=0}^{\infty}\int d\text{LIPS}_{n+2}\; \rho_A^{(n)}(\{p\},\{k\}),
\label{eqn-yfsmth-1}
\end{equation}
where $\text{LIPS}_{n+2}$ denotes Lorentz-invariant phase space for $n+2$ particles, $A=\text{CEEX},\;\text{EEX}$, the incoming and outgoing fermion momenta are abbreviated as $\{p\}$ and the $n$ photon momenta are denoted by $\{k\}$.
Thanks to the use of conformal symmetry, full $2+n$ body  phase space is covered without any
approximations. The respective MC algorithm's details are covered in Ref.~\cite{Jadach:1999vf}.
From Refs.~\cite{Jadach:2000ir,Jadach:1999vf} we have that 
\begin{equation}
\rho_{\text{CEEX}}^{(n)}(\{p\},\{k\})=\frac{1}{n!}e^{Y(\Omega;\{p\})}\bar{\Theta}(\Omega)\frac{1}{4}\sum_{\text{helicities}\;{\{\lambda\},\{\mu\}}}
\left|\Meu\left(\st^{\{p\}}_{\{\lambda\}}\st^{\{k\}}_{\{\mu\}}\right)\right|^2.
\label{eqn-yfsmth-2}
\end{equation}
(See Refs.~\cite{Jadach:2000ir,Jadach:1999vf} for the corresponding formula for the $A=\text{EEX}$ case.) Here,  $Y(\Omega;\{p\})$ is the YFS infrared exponent.
The respective infrared integration limits are specified by the region $\Omega$ and its characteristic function
$\Theta(\Omega,k)$ for a photon of energy $k$, with $\bar\Theta(\Omega;k)=1-\Theta(\Omega,k)$ and $$\bar\Theta(\Omega)=\prod_{i=1}^{n}\bar\Theta(\Omega,k_i).$$
By definition, $\Theta(\Omega,k)=1$  for $k\in\Omega$ 
and $\Theta(\Omega,k)=0$ for $k\not\in \Omega$.
As we will need it in what follows, we note that for  $\Omega$ defined with the condition $k^0<E_{\min},$ the YFS infrared exponent reads
\begin{equation}
  \label{eq:YFS-ffactor}
  \begin{split}
   &Y(\Omega;p_1,...,p_4)
  =   Q_e^2   Y_\Omega(p_1,p_2)  +Q_f^2   Y_\Omega(p_3,p_4)\\
&\qquad\qquad
     +Q_e Q_f Y_\Omega(p_1,p_3)  +Q_e Q_f Y_\Omega(p_2,p_4) 
     -Q_e Q_f Y_\Omega(p_1,p_4)  -Q_e Q_f Y_\Omega(p_2,p_3).
  \end{split}
\end{equation}
where 
\begin{equation}
\label{eq:form-factor}
Y_\Omega(p,q) 
     \equiv  2 \alpha \tilde{B}(\Omega,p,q)   +2 \alpha \Re B(p,q) \\
\end{equation}
is a sum of the real infrared contribution determined by the real emission infrared function $\tilde{B}$ and the virtual infrared contribution determined by the virtual infrared function $B$. The latter two infrared functions are given by
\begin{equation}
\label{eq:infrared-fns}
\begin{split}
     \tilde{B}(\Omega,p,q)\equiv & -\;{ 1 \over 8\pi^2} \int {d^3k\over k^0} \Theta(\Omega;k)
                       \bigg({p\over kp} - {q\over kq} \bigg)^2 ,\\
      B(p,q)  \equiv    &  \int {d^4k\over {k^2-\lambda^2}+i\epsilon} {i\over (2\pi)^3} 
                       \bigg( {2p-k \over -i\epsilon+2kp-k^2} - {2q+k \over i\epsilon+2kq+k^2} \bigg)^2,
\end{split}
\end{equation}
where $\lambda \downarrow 0$ is a photon-mass infrared regulator.
See Refs.~\cite{Jadach:1999vf,Jadach:2000ir,Jadach:2013aha} for the definitions of the CEEX amplitudes~$\{\Meu\}$ and
Ref.~\cite{Jadach:2022vf} for their most recent implementation and application in the C++ version of 5.0 \KK{MC} now denoted as \KK{MC-ee}\footnote{This notation distinguishes it from the MC \KK{MC-hh}~\cite{kkmchh,kkmchh1,kkmchh2} which calculates CEEX ${\cal O}(\alpha^2 L^2, \alpha^2L)$ EW corrections to the Drell-Yan single $Z/\gamma*$ production processes with decay to lepton pairs.}.
As the respective implementation is described in Ref.~\cite{Jadach:2000ir}, we do not repeat it here.
In {\KK}MC-ee, the CEEX amplitudes $\{\Meu\}$ in
Eqs.~(\ref{eqn-yfsmth-1},\ref{eqn-yfsmth-2}) are exact 
in ${\cal O}(\alpha^2 L^2, \alpha^2L)$ in the sense that all terms in the respective cross section at orders ${\cal O}(\alpha^0),\;{\cal O}(\alpha),\; {\cal O}(\alpha L),\;{\cal O}(\alpha^2 L), \; \text{and}\; {\cal O}(\alpha^2 L^2)$ are all included in our result for that cross section via the corresponding hard photon residuals $\hat\beta_n$. Here the big log is $L=\ln\frac{Q^2}{m^2}$ where
$Q$ is the respective hard 4-momentum transfer and the charged lepton masses and the quark masses determine $m$, depending on the specific process under consideration. We follow Ref.~\cite{mstw-mass} and use the current quark masses~\cite{PDG:2016} $m_u = 2.2 \text{MeV}, \; m_d = 4.7 \text{MeV}, \;  m_s = 0.150\text{GeV}, \; m_c = 1.2 \text{GeV},\;  m_b = 4.6\text{GeV} \text{and}\; m_t = 173.5 \text{GeV}$\footnote{See Ref.~\cite{kkmchh1} for a relevant discussion of the uncertainty of our results due to realistic uncertainties on our values of the current quark masses - we find a shift of our effects at the level of 10\% of the effects themselves due to the latter uncertainties.}. We note for completeness that in our MC's all real and soft virtual photonic corrections have $\alpha=\alpha(0)=\frac{1}{137.035999...}$, since real photons are massless.
For hard virtual QED corrections, we use $\alpha=\alpha(Q)\equiv=\alpha(0)/(1-\Delta\alpha(Q)),$ with the vacuum polarization $\Delta\alpha(Q)$ taken after Refs.~\cite{zfitter1,zfitter6:1999,zfitter:2006,dizet642,Arbuzov:2020coe} using the hadronic contribution from Ref.~\cite{fjeger-fccwksp2019} in an on-shell renormalization scheme with $\alpha(0), G_\mu,$ and $M_Z$ as inputs -- see for example Refs.~\cite{zfitter:2006,fjeger-fccwksp2019}. For completeness, we note that the EEX realization in {\KK}MC-ee includes as well the exact ${\cal O}(\alpha^3 L^3)$ corrections. The user always has the option to switch on this correction as needed. Here,
we explore collinear enhancement of the respective YFS resummation algebra while maintaining the level of exactness just described.\par
\section{Collinearly Enhanced YFS Theory}
The fundamental idea of the YFS resummation is to isolate and resum to all orders in $\alpha$ the infrared singularities so that these singularities are canceled to all such orders between real and virtual corrections. The question naturally arises as to what, if any, non-soft\footnote{The soft collinear singularities are already included since the YFS soft limit is exact.} collinear singularities are also resummed in the YFS resummation algebra. We consider virtual and real corrections in turn.\par
\subsection{Virtual Corrections} 
In the case of virtual corrections we can see the answer to this question by examining the exact result for the YFS IR exponent $Y(\Omega;\{p\})$ in eq.(\ref{eqn-yfsmth-2}): focusing on the $s$-channel and $s'$-channel contributions for reasons of pedagogy, we have,
dropping terms of ${\cal O}(m^2/s,\;m^2/s' )\; \text{where} \; s=(p_1+p_2)^2, \;s'=(p_3+p_4)^2$, 
\begin{equation}
\label{eq:form-factor1}
\begin{split}
  Y_e(\Omega_I;p_1,p_2)
 &=   \gamma_e \ln {2E_{min}\over \sqrt{2p_1p_2}}     +{1\over 4}\gamma_e
      +Q_e^2 {\alpha\over\pi} \bigg( -{1\over 2} +{\pi^2\over 3}\bigg),\\
  Y_f(\Omega_F;p_3,p_4)
 &=   \gamma_f \ln {2E_{min}\over \sqrt{2p_3p_4}}     +{1\over 4}\gamma_f
      +Q_f^2 {\alpha\over\pi} \bigg( -{1\over 2} +{\pi^2\over 3}\bigg),
\end{split}
\end{equation}
with
\begin{equation}
\label{eq:form-factor2}
\gamma_e = 2 Q_e^2 {\alpha\over \pi} \bigg( \ln {2p_1p_2\over m_e^2} -1 \bigg),\quad
\gamma_f         = 2 Q_f^2 {\alpha\over \pi} \bigg( \ln {2p_3p_4\over m_f^2} -1 \bigg),
\end{equation}
using an obvious notation for the $f\bar{f}$ production process.
We see that the YFS exponent has also resummed the non-infrared collinear big log term $\frac{1}{2}Q^2{\alpha\over\pi} L$ to the infinite order in both the ISR and FSR contributions, where $Q=Q_e,\; Q_f,$ respectively. The question naturally arises as to whether or not the YFS algebra can be extended to resum further collinear big log contributions. Indeed, it is known~
\cite{gribv-lptv:1972} that from the QED form factor the term $\frac{3}{2}{\alpha\over\pi} L$ exponentiates. Does YFS algebra allow for this? Note that we are not abandoning the YFS approach for the collinear factorization approach. We are asking the very limited but important question of whether, within the YFS approach, the entire term $\frac{3}{2}{\alpha\over\pi} L$ can be shown to exponentiate as found by Ref.~\cite{gribv-lptv:1972} in the collinear factorization approach. 
\par

\begin{figure}[ht]
\begin{center}
\setlength{\unitlength}{1in}
\begin{picture}(6,2.4)(0,0)
\put(1.0,0.2){\includegraphics[width=6in]{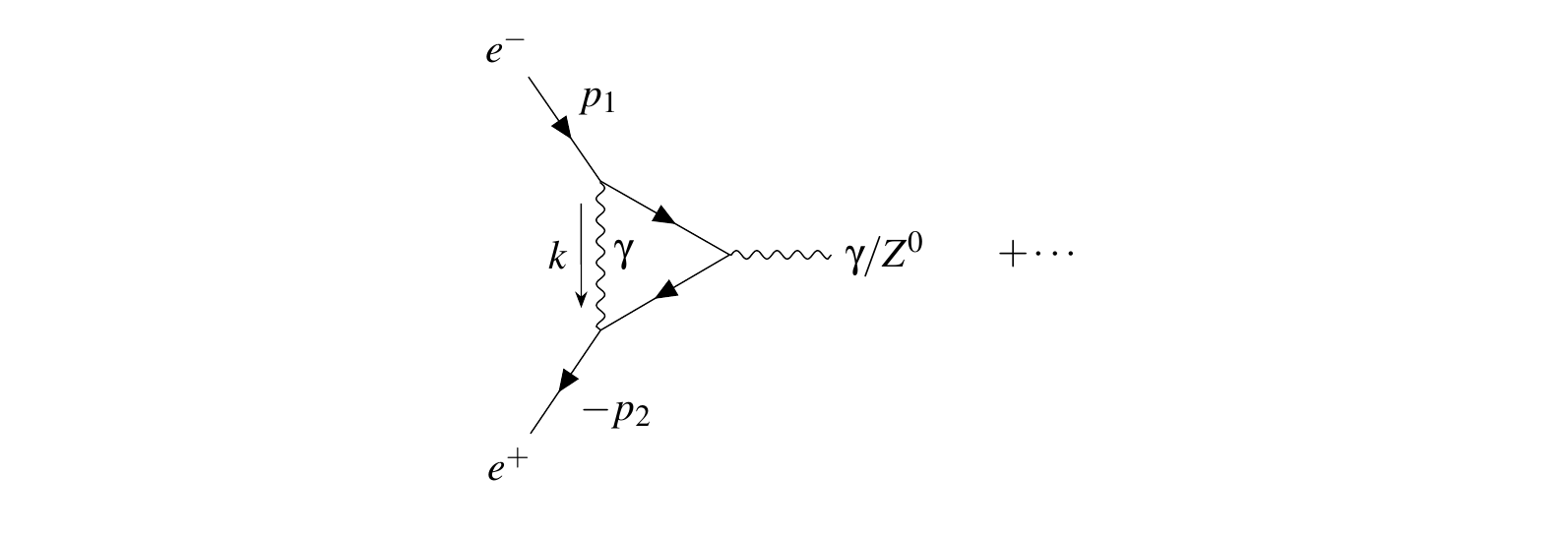}}
\end{picture}
\end{center}
\vspace{-10mm}
\caption{Virtual corrections which generate the YFS infrared function $B$.  Self-energy contributions are not shown.}
\label{fig1}
\end{figure}

To investigate this point, we focus on the derivation of the YFS form factor as illustrated in Fig. ~\ref{fig1}. 
This loop in which we have labeled the virtual photon momentum $k$ gives us the amplitude factor 
\begin{equation}
\begin{split}
{\cal M}_\mu &= \frac{\int d^4k}{(2\pi)^4}\frac{-i}{k^2+i\epsilon}\bar{v}(p_2)(-iQ_ee)\gamma^\alpha \frac{i}{-\!\not p_2-\!\not k-m+i\epsilon}(-ie)\gamma_\mu(v_A-a_A\gamma_5)\\
                      & \quad \quad \quad \quad \frac{i}{\!\not p_1-\!\not k-m+i\epsilon}(-iQ_ee)\gamma_\alpha u(p_1)
\end{split}
\label{yfsalg1}
\end{equation}
where $A=\gamma\;\text{or}\; Z$ according to whether we have the photon or Z exchange in the s-channel, respectively.
The couplings $v_A,\;a_A$ at the Standard Theory EW couplings in units of the positron charge $e$.
Scalarising the fermion propagator denominators gives us
\begin{equation}
{\cal M}_\mu = -ie\frac{\int d^4k(-i{Q_e}^2e^2)}{(2\pi)^4}\frac{1}{k^2+i\epsilon}\bar{v}(p_2)\gamma^\alpha \frac{-\!\not p_2-\!\not k+m}{k^2+2kp_2+i\epsilon}\gamma_\mu(v_A-a_A\gamma_5)\frac{\!\not p_1-\!\not k-m}{k^2-2kp_1+i\epsilon}\gamma_\alpha u(p_1).
\label{yfsalg2}
\end{equation}
The numerator factors for the fermion propagators in multiplication with the respective gamma matrices can be re-written, using the equations of motion, as
\begin{equation}
\begin{split}
(\!\not p_1-\!\not k-m)\gamma_\alpha u(p_1)&=\{(2p_1- k)_\alpha-\frac{1}{2}[\!\not k,\gamma_\alpha]\}u(p_1), \quad\qquad (a)\\
\bar{v}(p_2)\gamma^\alpha(-\!\not p_2-\!\not k+m)&=\bar{v}(p_2)\{-(2p_2+k)^\alpha+\frac{1}{2}[\!\not k,\gamma^\alpha]\}, \;\qquad (b).
\end{split}
\label{yfsalg3}
\end{equation}
This allows us to identify the contribution to $2Q_e^2\alpha B(p_1,p_2)$ corresponding to the cross-term in the virtual IR function on the RHS of eq.(\ref{eq:form-factor}):
\begin{equation}
2Q_e^2\alpha B(p_1,p_2)|_{\text{cross-term}}=\frac{\int d^4k(i{Q_e}^2e^2)}{8\pi^4}\frac{1}{k^2+i\epsilon}\frac{(2p_1-k)(2p_2 +k)}{(k^2-2kp_1+i\epsilon)(k^2+2kp_2+i\epsilon)}.
\label{cross-term}
\end{equation}
This term, together with the two squared terms in $2\alpha Q_e^2B(p_1,p_2),$ leads to the exponentiation of $\frac{1}{2}Q_e^2{\alpha\over\pi} L$ as we have indicated. 

If we look at the two commutator terms on the RHS of eq.(\ref{yfsalg3}), we see that, instead of dropping them from the YFS algebra as it is usually done, we can analyze them further for a possible IR finite collinearly enhanced improvement of the YFS virtual IR 
function $B$. For, if we isolate the collinear parts of $k$ via the change of variables~\cite{sudakov-1956}
\begin{equation}
   k = c_1p_1 +c_2p_2+k_\perp
\label{yfsalg4}
\end{equation}
where $p_1k_\perp=0=p_2k_\perp$,
we have the relations
\begin{equation}
\begin{split}
c_1 &= \frac{p_1p_2}{(p_1p_2)^2-m^4}p_2k - \frac{m^2}{(p_1p_2)^2-m^4}p_1k \xrightarrow[CL]{}\frac{p_2k}{p_1p_2} \\
c_2 &= \frac{p_1p_2}{(p_1p_2)^2-m^4}p_1k - \frac{m^2}{(p_1p_2)^2-m^4}p_2k \xrightarrow[CL]{}\frac{p_1k}{p_1p_2},
\end{split}
\label{yfsalg5}
\end{equation}
where $CL$ denotes the collinear limit in which terms ${\cal O}(m^2/s)$ are dropped. Here $s$ is the center-of-mass\footnote{The center-of-mass system is the system in which the total 3-momentum is zero.} (cms) squared energy
and, as already noted, satisfies $s=2p_1p_2+2m^2\simeq 2p_1p_2$.  This means that the numerator term $(2p_1-k)^\alpha$ in eq.(\ref{yfsalg3}(a)) combines with the commutator term in the eq.(\ref{yfsalg3}(b)) to produce the
the numerator contribution 
\begin{equation}
\begin{split}
\bar{v}(p_2)\{(2p_1-k)_\alpha\frac{1}{2}[\!\not k,\gamma^\alpha]\}\gamma_\mu(v_A-a_A\gamma_5)u(p_1)&=\bar{v}(p_2)[\!\not k,\!\not p_1]\gamma_\mu(v_A-a_A\gamma_5)u(p_1)\\
                                                                                                 &\xrightarrow[CL]{}\bar{v}(p_2)[c_2\!\not p_2 ,\!\not p_1]\gamma_\mu(v_A-a_A\gamma_5)u(p_1)\\
                                                                                                 &\xrightarrow[CL]{}\bar{v}(p_2)(-2c_2p_1p_2)\gamma_\mu(v_A-a_A\gamma_5)u(p_1) \\
                                                                                                 &\xrightarrow[CL]{}\bar{v}(p_2)(-2p_1k)\gamma_\mu(v_A-a_A\gamma_5)u(p_1). \\
\end{split}
\label{yfsalg6}
\end{equation}
Similarly, the numerator term $-(2p_2+k)^\alpha$ in eq.(\ref{yfsalg3} (b)) combines with the commutator term in eq.(\ref{yfsalg3}(a)) to produce the numerator contribution
\begin{equation}
\begin{split}
\bar{v}(p_2)\gamma_\mu(v_A-a_A\gamma_5)\{-(2p_2+ k)^\alpha(-\frac{1}{2}[\!\not k,\gamma_\alpha])\}u(p_1)&=\bar{v}(p_2)\gamma_\mu(v_A-a_A\gamma_5)[\!\not k,\!\not p_2]u(p_1)\\
                                                                                       &\xrightarrow[CL]{}\bar{v}(p_2)\gamma_\mu(v_A-a_A\gamma_5)[c_1\!\not p_1,\!\not p_2]u(p_1)\\
                                                                                       &\xrightarrow[CL]{}\bar{v}(p_2)\gamma_\mu(v_A-a_A\gamma_5)(2c_1p_1p_2)u(p_1)\\
                                                                                       &\xrightarrow[CL]{}\bar{v}(p_2)\gamma_\mu(v_A-a_A\gamma_5)(2p_2k)u(p_1).\\
\end{split}
\label{yfsalg7}
\end{equation}
We therefore have the shift of the factor $(2p_1-k)(2p_2+k)$ on the RHS of eq.(\ref{cross-term}) as
\begin{equation}
(2p_1-k)(2p_2+k)\xrightarrow[CL]{} (2p_1-k)(2p_2+k) +2p_1k -2p_2k.
\label{yfsalg8}
\end{equation} 

The term in the numerator which is quadratic in the commutator ($C^2$) is superficially divergent in the UV so that we cannot drop
$k_\perp$ naively. Instead of doing that, we proceed directly: {\small
\begin{equation}
2Q_e^2\alpha B(p_1,p_2)|_{C^2}{\cal M}_{B\mu}\equiv \int d^4k\frac{(i{Q_e}^2e^2)}{32\pi^4}\frac{1}{k^2+i\epsilon}\frac{\bar{v}(p_2)[\!\not k,\gamma^\alpha]\gamma_\mu[\!\not k,\gamma_\alpha](-ie)(v_A-a_A\gamma_5)u(p_1)}{(k^2-2kp_1+i\epsilon)(k^2+2kp_2+i\epsilon)}\bigg|_{CL'},
\label{quadratic-term}
\end{equation} }
where we define
\begin{equation}
{\cal M}_{B\mu} = -ie\bar{v}(p_2)\gamma_\mu(v_A-a_A\gamma_5)u(p_1)
\end{equation}
in an obvious notation and we restrict further the definition of $CL'$ here to include only contributions singular in the limit 
$m^2/s \rightarrow 0$ -- $CL'$ differs from $CL$ by constant terms that we drop here. There are four terms in the numerator on the RHS of eq.(\ref{quadratic-term}) from the respective sum of gamma matrix products $$\!\not k \gamma^\alpha \gamma_\mu \!\not k \gamma_\alpha -\!\not k \gamma^\alpha \gamma_\mu \gamma_\alpha \!\not k 
- \gamma^\alpha\!\!\!\!\not k \gamma_\mu \!\not k \gamma_\alpha + 
\gamma^\alpha\!\!\!\!\not k \gamma_\mu \gamma_\alpha \!\not k=\{\gamma^\lambda \gamma^\alpha \gamma_\mu \gamma^{\lambda'}\gamma_\alpha -\gamma^\lambda \gamma^\alpha \gamma_\mu \gamma_\alpha \gamma^{\lambda'}\; - \gamma^\alpha \gamma^\lambda \gamma_\mu \gamma^{\lambda'} \gamma_\alpha $$
$$ \quad \qquad \qquad \qquad \qquad \qquad \qquad \qquad \qquad + \gamma^\alpha \gamma^\lambda \gamma_\mu \gamma_\alpha \gamma^{\lambda'}\}k_\lambda k_{\lambda'}\equiv N^{\lambda\lambda'}_\mu k_\lambda k_{\lambda'},$$ where this latter equivalence sign serves to define $N^{\lambda\lambda'}_\mu$. Using n-dimensional methods~\cite{BW1a}, we see
that, when we combine the denominators to get the standard Feynman parametrization, we need to evaluate {\small
\begin{equation}
I_\mu= 2\int_0^1d\alpha_1\int_0^{1-\alpha_1}d\alpha_2 \int d^nk'\frac{(i{Q_e}^2e^2)}{32\pi^4}\frac{\bar{v}(p_2)N^{\lambda \lambda'}_\mu[\frac{{k'}^2}{n}g_{\lambda\lambda'}+\Delta_\lambda \Delta_{\lambda'}](-ie)(v_A-a_A\gamma_5)u(p_1)}{[{k'}^2-\Delta^2+i\epsilon]^3}\bigg|_{CL'},
\label{quadratic-term2}
\end{equation} }
where $\Delta=\alpha_1p_1-\alpha_2p_2$. From the equations of motion we see that the terms involving $\Delta$ on the RHS of eq.(\ref{quadratic-term2}) do not make a collinearly enhanced contribution. Computation of the term contracted with $g_{\lambda \lambda'}$ on the RHS of eq.(\ref{quadratic-term2}) gives us 
\begin{equation}
I_\mu={\bigg\{} \frac{ -3Q_e^2\alpha}{ 4\pi}{\cal M}_{B\mu}{\bigg\}}{\bigg|}_{CL'}\equiv 0
\end{equation}
so that there are no collinearly enhanced contributions from $I_\mu$. Eq.(\ref{yfsalg8}) gives the complete collinear enhancement of $B$.

This change in $B$ does not affect its IR behavior because the shifted terms are IR finite. Thus, the entire YFS IR resummation is unaffected. But, the shifted terms can be seen to extend the YFS IR exponentiation to obtain the entire exponentiated $\frac{3}{2}Q_e^2\alpha L$.

Specifically, we have
\begin{equation}
\begin{split}
2\alpha Q_e^2\Delta B(p_1,p_2) &=\frac{\int d^4k(iQ_e^2e^2)}{8\pi^4}\frac{1}{k^2+i\epsilon}\frac{2p_1k - 2p_2k}{(k^2-2kp_1+i\epsilon)(k^2+2kp_2+i\epsilon)}\\
                                     &=2\int_{x_i\ge0,i=1,2,3} d^3x\delta(1-x_1-x_2-x_3)\frac{\int d^4k'(iQ_e^2e^2)}{8\pi^4}\frac{2(p_1 - p_2)p_x}{(k'^2-d+i\epsilon)^3}
\end{split}                                     
\label{yfsalg9}
\end{equation}
where $d=p_x^2$ with $p_x=x_1p_1-x_2p_2$. We get
\begin{equation}
2Q_e^2\alpha \Re \Delta B(p_1,p_2) = Q_e^2{\alpha\over\pi} L.
\label{yfsalg10}
\end{equation}
We see that indeed the entire term ${3\over2}Q_e^2{\alpha\over\pi} L$ is now exponentiated by our collinearly improved YFS virtual IR function $B_{CL}$ given by 
\begin{equation}
\begin{split}
B_{CL}&=B+\Delta B\\
          &= \int {d^4k\over k^2} {i\over (2\pi)^3} 
                       \bigg[\bigg( {2p-k \over 2kp-k^2} - {2q+k \over 2kq+k^2} \bigg)^2-\frac{4pk-4qk}{(2pk-k^2)(2qk+k^2)}\bigg].
\end{split}
\end{equation}
We stress again that the YFS IR resummation calculus is unaffected by the use of $B_{CL}$ instead of $B$\footnote{In Ref.~\cite{Jadach-unpublhd-2002} one of us (SJ) has identified the integrated form of $B_{CL}$ by matching with the respective Sudakov form factor.}. Indeed, the entire algebra of the YFS resummation is unaffected by the use of $B_{CL}$ instead of $B$: for example, eqs.(2.2) from Ref.~\cite{yfs:1961} become, for the case of virtual photon corrections to an exact amplitude $M$,
\begin{equation}
\begin{split}
M_0&=\mathfrak{m}_0 \\
M_1 &= \mathfrak{m}_0\alpha B_{CL}+\mathfrak{m}_1,\\
M_2 &= \mathfrak{m}_0 \frac{(\alpha B_{CL})^2}{2}+\mathfrak{m}_1\alpha B_{CL}+\mathfrak{m}_2\\
       & \vdots\\
M_n &=\sum_{r=0}^{n} \mathfrak{m}_{n-r}\frac{(\alpha B_{CL})^r}{r!}\Rightarrow\\
M & = \sum_{n=0}^\infty M_n = e^{\alpha B_{CL}}\sum_{n=0}^\infty \mathfrak{m}_n,
\end{split}
\end{equation} 
so that the YFS algebra guarantees there is no double counting when we use $B_{CL}$ instead of $B$. Here, $M_n$ is the {\it exact} amplitude with n virtual photons and $\mathfrak{m}_n$ the corresponding {\it infrared finite} residual. The same algebra leads to the exponentiation of $B_{CL}$ just as it does for $B$. We see that the infrared finite residuals $\mathfrak{m}_n$ are shifted by infrared finite terms when we use $B_{CL}$ instead of $B$. This leads to improved precision for a given level of exactness in the $M_n$. The corresponding improved YFS form factors are now in the $s$ and $s'$ channels
\begin{equation}
\begin{split}
  Y_{CL,e}(\Omega_I;p_1,p_2)
 &=   \gamma_e \ln {2E_{min}\over \sqrt{2p_1p_2}}     +{3\over 4}\gamma_e
      +Q_e^2 {\alpha\over\pi} \bigg( {1\over 2} +{\pi^2\over 3}\bigg),\\
  Y_{CL,f}(\Omega_F;p_3,p_4)
 &=   \gamma_f \ln {2E_{min}\over \sqrt{2p_3p_4}}     +{3\over 4}\gamma_f
      +Q_f^2 {\alpha\over\pi} \bigg( {1\over 2} +{\pi^2\over 3}\bigg),
\end{split}
\end{equation}
using an obvious notation.
\par
\subsection{Real Corrections}
\begin{figure}[ht]
\begin{center}
\setlength{\unitlength}{1in}
\begin{picture}(6,2.4)(0,0)
\put(0.1,0.2){\includegraphics[width=6in]{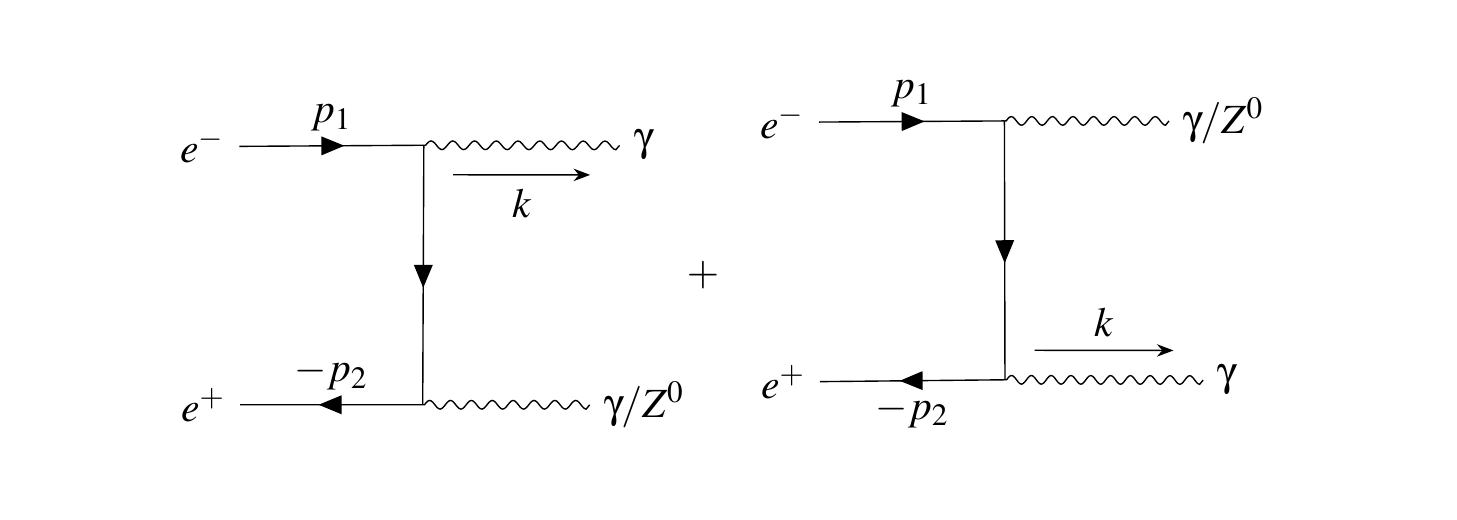}}
\end{picture}
\end{center}
\vspace{-10mm}
\caption{Real corrections which generate the YFS infrared function $\tilde{B}$.}
\label{fig2}
\end{figure}

In some applications, it can also be of interest to improve collinearly the YFS real emission IR resummation algebra. In this connection we recall that the original YFS EEX formulation of the respective algebra leads to the
formula for the YFS IR function $\tilde{B}$ given above in eq.(\ref{eq:form-factor}). The situation is 
illustrated in Fig.~\ref{fig2}.
Following the steps in the usual YFS algebra for real emission, we see that the corresponding contribution to $2\alpha \tilde{B}$ can be isolated from the respective squared matrix element integrated over the corresponding photon phase space as follows:
\begin{equation}
\begin{split}
2\alpha Q_e^2\tilde{B}{\cal M}^\dagger_{B\mu}{\cal M}_{B\mu'} &=\frac{\int d^3k (-1)e^2 Q_e^2}{2k_0(2\pi)^3}\bigg[\frac{\bar{u}(p_1)\large(2p_1^\lambda - k^\lambda + \frac{1}{2}[\!\not k,\gamma^\lambda]\large)\gamma_\mu(v_A-a_A\gamma_5)v(p_2)}{k^2-2kp_1}\\
                                              & + \frac{\bar{u}(p_1)\gamma_\mu(v_A-a_A\gamma_5)\large(-2p_2^\lambda + k^\lambda + \frac{1}{2}[\!\not k,\gamma^\lambda]\large)v(p_2)}{k^2-2kp_2}\bigg]\\
                                              & \bigg[\frac{\bar{v}(p_2)\gamma_{\mu'}(v_A-a_A\gamma_5)\large(2p_{1\lambda} - k_\lambda - \frac{1}{2}[\!\not k,\gamma_\lambda]\large)u(p_1)}{k^2-2kp_1}\\
                                             & + \frac{\bar{v}(p_2)\large(-2p_{2\lambda} + k_\lambda - \frac{1}{2}[\!\not k,\gamma_\lambda]\large)\gamma_{\mu'}(v_A-a_A\gamma_5)u(p_1)}{k^2-2kp_2}\bigg]\bigg|_{k^2=0} + K_{\mu\mu'}
\end{split}
\label{eq-real1}
\end{equation}
where $K_{\mu\mu'}$ is infrared finite. We can immediately see that if we drop the commutator terms on the RHS of eq.(\ref{eq-real1}) we recover the usual YFS formula for $2\alpha Q_e^2\tilde{B}$. We can again isolate collinearly enhanced contributions by using the representation in eq.(\ref{yfsalg4}) for $k$. However, we have to respect the condition $k^2=0.$ This means
that we have to maintain $0=(c_1^2+c_2^2)m^2+ 2c_1c_2p_1p_2-|k_\perp|^2$. Upon isolating the collinearly enhanced contributions subject to this latter constraint we get the following collinear enhancement of $\tilde{B}$:
\begin{equation}
\begin{split}
2\alpha Q_e^2\tilde{B}_{CL} &=\frac{-\alpha Q_e^2}{4\pi^2}\int\frac{d^3k}{k_0}\bigg\{\large(\frac{p_1}{kp_1} - \frac{p_2}{kp_2}\large)^2 +\frac{1}{kp_1}\large(2 -\frac{kp_2}{p_1p_2}\large)\\
                                          &\qquad\qquad+\frac{1}{kp_2}\large(2 -\frac{kp_1}{p_1p_2}\large)\bigg\}.
\end{split}
\label{eq-real2}
\end{equation}
The extra non-IR divergent terms can be seen to reproduce the respective collinear big log of the exact result in Ref.~\cite{berends-neerver-burgers:1988} in the soft regime. 
Specifically, if we integrate over the new collinearly enhanced terms we get the contribution
\begin{equation}
\begin{split}
2\alpha Q_e^2\Delta\tilde{B} &=\frac{-\alpha Q_e^2}{4\pi^2}\int^{k_0\le E_{min}}\frac{d^3k}{k_0}\bigg\{\frac{1}{kp_1}\large(2 -\frac{kp_2}{p_1p_2}\large) + \frac{1}{kp_2}\large(2 -\frac{kp_1}{p_1p_2}\large)\bigg\}\\
                                          &= \frac{\alpha}{\pi}Q_e^2\large(-2v_{min}L+\frac{1}{2}v_{min}^2(L-1)\large)
\end{split}
\label{eq-real2a}
\end{equation}
where $v_{min}=2E_{min}/\sqrt{s}$, in agreement with Ref.~\cite{berends-neerver-burgers:1988}\footnote{The linear term in $v_{min}$ on the RHS of the second line in eq.(\ref{eq-real2a}) would have the coefficient $L-1$ instead of $L$ if we would keep the terms of ${\cal O}(m^2/s)$ in the $c_i$.}.
This then leads us to consider the realization of eq.(\ref{eq-real2}) in the CEEX formalism. To this we now turn.\par
For the CEEX formalism, we revisit Fig.~\ref{fig2} with the use of amplitude-level isolation of real IR divergences.
We follow Ref.~\cite{ceex2:1999} and introduce the Kleiss-Stirling~\cite{kleiss-stirling:1985} photon polarization vectors 
so that the amplitude in Fig. 2 can be written as, for the photon polarization $\sigma$ and $e^-$  helicity $\sigma'$,
\begin{equation}
{\cal M}_\mu = {\cal M}_{B\mu}\sfac_{CL,\sigma}(k),
\label{eq-real3}
\end{equation}
where we define the collinearly enhanced soft (eikonal) amplitude factor, an extension of the corresponding factor defined in Ref.~\cite{ceex2:1999}, via
\begin{equation}
\begin{split}
\sfac_{CL,\sigma}(k) = \sqrt{2}Q_ee\bigg[-\sqrt{\frac{p_1\zeta}{k\zeta}}\frac{<k\sigma|\hat{p}_1 -\sigma>}{2p_1k}
    + \delta_{\sigma'\;-\sigma}\sqrt{\frac{k\zeta}{p_1\zeta}}\frac{<k\sigma|\hat{p}_1 \sigma'>}{2p_1k}\\
    + \sqrt{\frac{p_2\zeta}{k\zeta}}\frac{<k\sigma|\hat{p}_2 -\sigma>}{2p_2k}+\delta_{\sigma' \sigma}\sqrt{\frac{k\zeta}{p_2\zeta}}\frac{<\hat{p}_2 \sigma'|k -\sigma>}{2p_2k}\bigg].
\end{split}
\label{eq-real4}
\end{equation}
We have introduced from Ref.~\cite{ceex2:1999} the notations
$\zeta\equiv (1,1,0,0)$ for our choice for the respective auxiliary vector in our Global Positioning of Spin (GPS)~\cite{gps:1998} spinor conventions with the consequent definition $\hat{p}= p - \zeta m^2/(2\zeta p)$
for any four vector $p$ with $p^2 = m^2.$ If we take the modulus squared of $\sfac_{CL,\sigma}(k)$ and sum over the respective photon helicities, we get
\begin{equation}
\begin{split}
\sum_{\sigma=\pm}\big|\sfac_{CL,\sigma}(k)\big|^2 &= 2Q_e^2 e^2\sum_{\sigma=\pm}\big[ (\frac{p_1\zeta}{k\zeta}+\frac{k\zeta}{2p_1\zeta}-1)\frac{k\hat{p}_1}{(kp_1)^2}+(\frac{p_2\zeta}{k\zeta}+\frac{k\zeta}{2p_2\zeta}-1)\frac{k\hat{p}_2}{(kp_2)^2}\\
&\qquad+\frac{1}{2kp_1\;kp_2}\big(\frac{1}{p_2\zeta}+\frac{1}{p_1\zeta}-\frac{2}{k\zeta}\big)(\hat{p}_1\zeta\;\hat{p}_2k+\hat{p}_1k\;\hat{p}_2\zeta-k\zeta\;\hat{p}_1\hat{p}_2)\big]\\
&\underset{CL'}{=} 2Q_e^2 e^2\big[-\frac{1}{2}\big(\frac{m^2}{(kp_1)^2}+\frac{m^2}{(kp_2)^2}-\frac{2p_1p_2}{kp_1\;kp_2}\big)-(\frac{1}{kp_1}+\frac{1}{kp_2})\\
&\qquad+\frac{kp_2}{2kp_1\;p_1p_2}+\frac{kp_1}{2kp_2\;p_1p_2}\big]\\
&=-Q_e^2 e^2\bigg\{\large(\frac{p_1}{kp_1} - \frac{p_2}{kp_2}\large)^2 +\frac{1}{kp_1}\large(2-\frac{kp_2}{p_1p_2}\large)+\frac{1}{kp_2}\large(2-\frac{kp_1}{p_1p_2}\large)\bigg\},
\end{split}
\label{eq-real5}
\end{equation}
in agreement with the integrand in eq.(\ref{eq-real2}) when we include the phase space factor $d^3k/(16\pi^3k_0)$ from the standard methods\footnote{We take the $z$-direction as that of $\vec{p}_1$ for definiteness.}. Thus, we see that the extra non-IR divergent contributions reproduce the known collinear big log contribution which is missed by the usual YFS algebra.\par

We have analyzed the s-channel terms in Eqs.(\ref{eq:YFS-ffactor},\ref{eq:form-factor}), which are evaluated explicitly in Eqs.(\ref{eq:form-factor1},\ref{eq:form-factor2}). As usual
with the YFS algebra, the extension to the remaining terms in  eq.(\ref{eq:YFS-ffactor}) is obtained by the respective substitutions exhibited therein, such as 
$(Q_e^2, p_a,\; p_b) \rightarrow (Q_e Q_f, p_a,\; p_c)$ to obtain the result for the t-channel, etc. We stress that we have only isolated the respective part of the attendant collinearly singular contributions which exponentiate whereas the arguments in Refs.~\cite{amati-pet-venez-1978,amati-pet-venez-1978-2,ellis-1979,cur-furm-pet-1980,c-s-s-1988,bertone-2019,bertone-2022}
isolate all collinear singularities.\par

\section{Summary}
In conclusion, we have extended the original YFS algebra to include collinear non-IR big logs that are known to be missed by the usual YFS soft functions. This implies that these new, collinearly enhanced soft functions will yield a higher level of accuracy for a given level of exactness in the IR-finite YFS hard photon residuals. They thus enhance the
set of the tools available to extend the CEEX YFS MC method to the other important processes in the future $e^+e^-$ colliders' precision physics programs in the effort to reach the new required precisions for these processes.

\vskip 2 mm
\centerline{\bf Acknowledgments}
\vskip 2 mm

The authors thank Prof. G. Giudice for the support 
and kind hospitality of the CERN TH Department. 
S.J. acknowledges funding from the European Union’s Horizon 2020 research and innovation programme under under grant agreement No 951754 and support of the National Science Centre, Poland, Grant No. 2019/34/E/ST2/00457.

\bibliography{Tauola_interface_design}{}
\bibliographystyle{utphys_spires}

\end{document}